\documentclass{elsart5p}

\usepackage{epsfig}

\usepackage{amssymb}

\usepackage{xspace}

\def\cref#1{Chapt.\,\ref{#1}}
\def\Cref#1{Chapter~\ref{#1}}
\def\sref#1{Sect.\,\ref{#1}}

\def\fref#1{Fig.\,\ref{#1}}

\def\1{\footnotemark[1]}
\def\and{\& }
\def\Cerenkov{\v{C}erenkov\xspace}

\def\deg{$^\circ$\xspace}

\def\gcm2{g/cm$^2$\xspace}

\def\Xmax{$X_{max}$\xspace}
\def\vxB{$\vec{v}\times\vec{B}$\xspace}
\def\Lstar{LOPES$^{\rm STAR}$\xspace}

\def\Section#1{\section{#1}}

\begin{document}

\begin{frontmatter}



\title{Measurement of Radio Emission from Extensive Air Showers with LOPES}


\author[4]{J.R. H\"orandel},
\author[1]{W.D. Apel},
\author[2,14]{J.C.~Arteaga},
\author[3]{T.~Asch},
\author[1]{F.~Badea},
\author[4]{L.~B\"ahren},
\author[1]{K.~Bekk},
\author[5]{M.~Bertaina},
\author[6]{P.L.~Biermann},
\author[1,2]{J.~Bl\"umer},
\author[1]{H.~Bozdog},
\author[7]{I.M.~Brancus},
\author[8]{M.~Br\"uggemann},
\author[8]{P.~Buchholz},
\author[4]{S.~Buitink},
\author[5,9]{E.~Cantoni},
\author[5]{A.~Chiavassa},
\author[2]{F.~Cossavella},
\author[1]{K.~Daumiller},
\author[2,15]{V.~de Souza},
\author[5]{F.~Di~Pierro},
\author[1]{P.~Doll},
\author[1]{M.~Ender},
\author[1]{R.~Engel},
\author[4,10]{H.~Falcke},
\author[1]{M.~Finger},
\author[11]{D.~Fuhrmann},
\author[3]{H.~Gemmeke},
\author[9]{P.L.~Ghia},
\author[11]{R.~Glasstetter},
\author[8]{C.~Grupen},
\author[1]{A.~Haungs},
\author[1]{D.~Heck},
\author[4]{A.~Horneffer},
\author[1]{T.~Huege},
\author[1]{P.G.~Isar},
\author[11]{K.-H.~Kampert},
\author[2]{D.~Kang},
\author[8]{D.~Kickelbick},
\author[3]{O.~Kr\"omer},
\author[4]{J.~Kuijpers},
\author[4]{S.~Lafebre},
\author[1]{K.~Link},
\author[12]{P.~{\L}uczak},
\author[2]{M.~Ludwig},
\author[1]{H.J.~Mathes},
\author[1]{H.J.~Mayer},
\author[2]{M.~Melissas},
\author[7]{B.~Mitrica},
\author[9]{C.~Morello},
\author[5]{G.~Navarra},
\author[1]{S.~Nehls},
\author[4]{A.~Nigl},
\author[1]{J.~Oehlschl\"ager},
\author[8]{S.~Over},
\author[2]{N.~Palmieri},
\author[7]{M.~Petcu},
\author[1]{T.~Pierog},
\author[11]{J.~Rautenberg},
\author[1]{H.~Rebel},
\author[1]{M.~Roth},
\author[7]{A.~Saftoiu},
\author[1]{H.~Schieler},
\author[3]{A.~Schmidt},
\author[1]{F.~Schr\"oder},
\author[13]{O.~Sima},
\author[4,16]{K.~Singh},
\author[7]{G.~Toma},
\author[9]{G.C.~Trinchero},
\author[1]{H.~Ulrich},
\author[1]{A.~Weindl},
\author[1]{J.~Wochele},
\author[1]{M.~Wommer},
\author[12]{J.~Zabierowski},
\author[6]{J.A.~Zensus}

\address[4]{Radboud University Nijmegen, Department of Astrophysics, 
         P.O. Box 9010, 6500 GL Nijmegen, The Netherlands }
\address[1]{Institut f\"ur Kernphysik, Forschungszentrum Karlsruhe, Germany}
\address[2]{Institut f\"ur Experimentelle Kernphysik,
 Universit\"at Karlsruhe, Germany}
\address[3]{IPE, Forschungszentrum Karlsruhe, Germany
}
\address[5]{Dipartimento di Fisica Generale dell'Universit{\`a} di Torino, Italy}
\address[6]{Max-Planck-Institut f\"ur Radioastronomie
 Bonn, Germany}
\address[7]{National Institute of Physics and Nuclear
 Engineering, Bucharest, Romania}
\address[8]{Fachbereich Physik, Universit\"at Siegen,
 Germany}
\address[9]{Istituto di Fisica dello Spazio Interplan
etario, INAF Torino, Italy}
\address[10]{ASTRON, Dwingeloo, The Netherlands}
\address[11]{Fachbereich Physik, Universit\"at Wuppertal, Germany}
\address[12]{Soltan Institute for Nuclear Studies, Lodz, Poland}
\address[13]{Department of Physics, University of Bucharest, Bucharest, Romania}
\address[14]{\small now at: Universidad Michoacana, Morelia, Mexico}
\address[15]{\small now at: Universidade de S\~ao Paulo, Instituto de
             F\'{\i}sica de S\~ao Carlos, Brasil}
\address[16]{\small now at: KVI, University of Groningen, The Netherlands}

\begin{abstract}
A new method is explored to detect extensive air showers: the measurement of
radio waves emitted during the propagation of the electromagnetic shower
component in the magnetic field of the Earth.  Recent results of the pioneering
experiment LOPES are discussed.  It registers radio signals in the frequency
range between 40 and 80 MHz.  The intensity of the measured radio emission is
investigated as a function of different shower parameters, such as shower
energy, angle of incidence, and distance to shower axis.  In addition, new
antenna types are developed in the framework of \Lstar and new methods are
explored to realize a radio self-trigger algorithm in real time.
\end{abstract}

\journal{Nuclear Instruments and Methods A}
\end{frontmatter}

\Section{Introduction}

An intense branch of astroparticle physics is the study of high-energy cosmic
rays to reveal their origin, as well as their acceleration and propagation
mechanisms \cite{behreview,cospar06,wuerzburg}. At energies exceeding
$10^{14}$~eV cosmic rays are usually studied by indirect measurements --- the
investigation of extensive air showers initiated by cosmic particles in the
atmosphere. Different techniques are applied, like the measurements of particle
densities and energies at ground level, or the observation of \Cerenkov and
fluorescence light. An alternative technique has been recently revitalized ---
the detection of radio emission from extensive air showers at energies
exceeding $10^{16}$~eV.

\begin{figure}[t]
 \epsfig{file=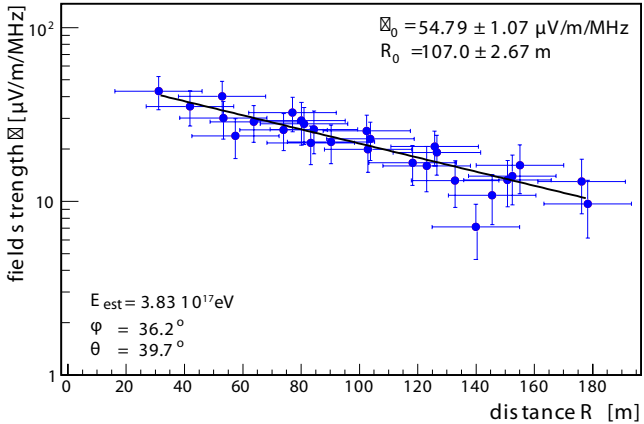,width=0.96\columnwidth}
 \caption{Measured field strength as a function of the distance to shower axis
          for an individual shower \cite{icrc09-nehls}.}
 \label{lat}	  
\end{figure}

Radio emission from air showers was experimentally discovered in 1965 at a
frequency of 44 MHz \cite{jelleynature}. The early activities in the 1960s and
1970s are summarized in \cite{allanrev}.  Only recently, fast analog-to-digital
converters and modern computer technology made a clear detection of radio
emission from air showers possible.  LOPES, a LOFAR Prototype Station had shown
that radio emission from air showers can be detected even in an environment
with relatively strong radio frequency interference (RFI) \cite{radionature}.
Further investigations of the radio emission followed with LOPES
\cite{badearadio,petrovicinclined,buitinkthunder,niglfreq,nigldirect} and the
CODALEMA experiment \cite{codalema,Ardouin:2006nb}, paving the way for this new
detection technique, see also \cite{Falcke:2008qk}. 

The LOPES experiment registers radio signals in the frequency range from 40 to
80~MHz \cite{lopesspie}. In this band are few strong man made radio
transmitters only, the emission from air showers is still strong (it decreases
with frequency), and background emission from the Galactic plane is still low.
An active short dipole has been chosen as antenna.  An inverted V-shaped dipole
is positioned about 1/4 of the shortest wavelength above an aluminum ground
plate. In this way a broad directional beam pattern is obtained.  LOPES
comprises 30 antennas \cite{nehlspune} located on site of the KASCADE-Grande
experiment \cite{kascadenim,grande} one of the best air shower experiments
operating in the energy range between $10^{14}$ and $10^{18}$~eV. The LOPES
data acquisition is triggered by large air showers registered with
KASCADE-Grande. The latter measures the showers simultaneously to LOPES and
delivers precise information on the shower parameters, such as shower energy as
well as position and inclination of the shower axis.  All antennas, including
the complete analog electronics chain, have been individually calibrated with a
reference radio source \cite{Nehls:2008ix}.

Most likely, the dominant emission mechanism of the radio waves in the
atmosphere is radiation due to the deflection of charged particles in the
Earth's magnetic field (geosynchrotron radiation)
\cite{huege2003,huege2005,huegefalcke}. 
In the frequency range of interest the wavelength of the radiation is large
compared to the size of the emission region: the typical thickness of the air
shower disc is about 1 to 2~m only. Thus, coherent emission is expected which
yields relatively strong signals at ground level.

\Section{Radio signal and shower parameters}

\begin{figure}[t]
 \epsfig{file=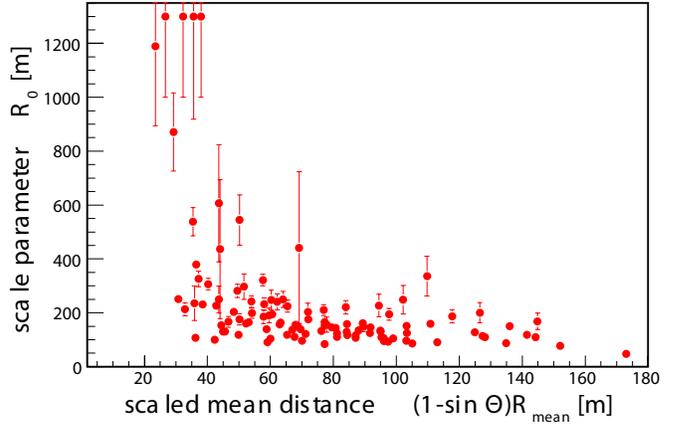,width=\columnwidth}
 \caption{Scale parameter $R_0$ as a function of the zenith angle times the
          mean distance to the shower axis \cite{icrc09-nehls}.}
 \label{latpar}	  
\end{figure}

One of the primary objectives of LOPES is to investigate the measured radio
signal as a function of parameters characterizing the extensive air shower.
For this purpose the shower parameters measured simultaneously with
KASCADE-Grande are irreplaceable.
An empirical relation has been found to express the expected east-west
component of the field strength at a distance $R$ from the shower axis as a
function of shower parameters \cite{horneffer-merida}
\begin{eqnarray}
\epsilon=(11\pm1) \left[ (1.16\pm0.025)-\cos \alpha \right] \cos \theta 
    \qquad\qquad\qquad \\
   \exp\left(-\frac{R}{(236\pm81)~{\rm m}}\right)
     \left(\frac{E_0}{10^{17}~{\rm eV}}\right)^{(0.95\pm0.04)}
     \left[\frac{\mu{\rm V}}{\rm m~MHz}\right]. \nonumber
\end{eqnarray}
$\alpha$ is the angle between the shower axis and the direction of the Earth
magnetic field (geomagnetic angle), $\theta$ the zenith angle of the shower,
and $E_0$ the energy of the shower inducing primary particle.  It is
interesting to note the absolute values of some parameters: the exponential
fall-off has a characteristic length of about 240~m, much larger than the
classical Moli\`ere radius of electrons in air ($\approx80$~m). This indicates
that the lateral distribution of the radio component is flatter as compared to
the electromagnetic component, an important fact to build large-scale radio
arrays with an economic antenna density.  The measured radio signal is almost
directly proportional to the shower energy ($[0.95\pm0.04]\approx1$). Such a
behavior is expected for a coherent emission of the radio waves from the air
showers.  This calibration of the measured radio signal is a first important
step towards the application of the radio detection as independent method to
register extensive air showers.

\begin{figure*}[t]
 \epsfig{file=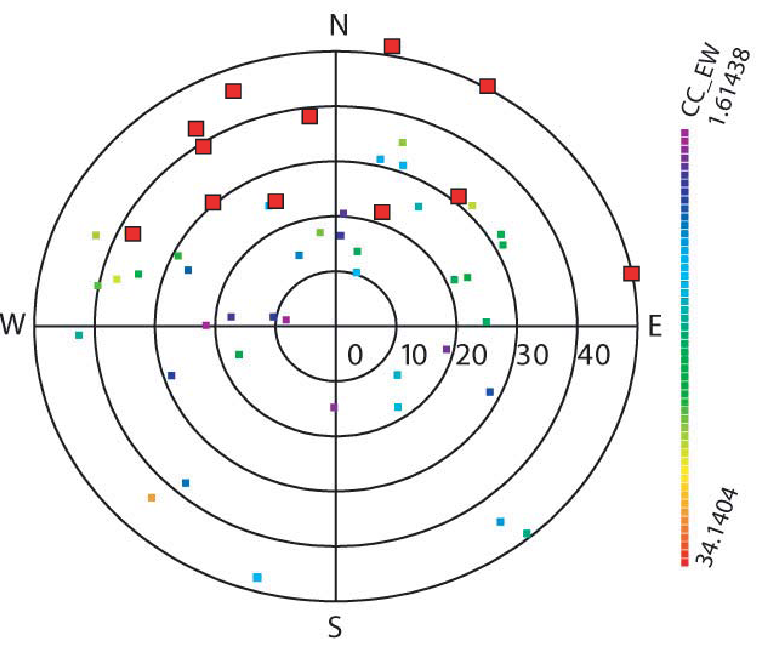,width=0.96\columnwidth}\hspace*{\fill}
 \epsfig{file=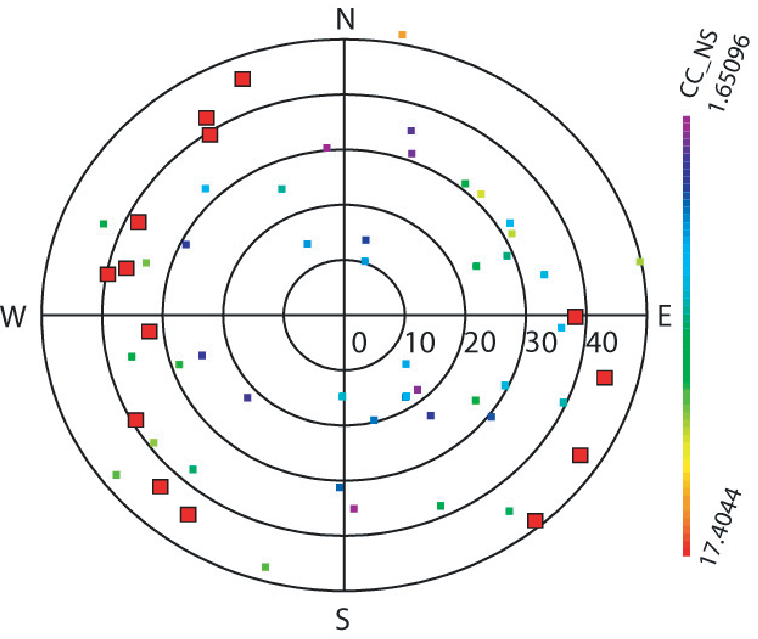,width=\columnwidth}
 \caption{Sky map of air showers detected in east-west (left) and north-south
	  (right) polarization direction \cite{isar-nim2009}. The boxes mark
	  the arrival direction of the air showers with the strongest radio
          signal.}
 \label{polsky}	  
\end{figure*}

Recently, the lateral distribution of the measured radio signals has been
investigated in more detail \cite{icrc09-nehls}.  A function of the form
\begin{equation}
  \epsilon = \epsilon_0 \exp\left(-\frac{R}{R_0}\right)
\end{equation}
has been fitted to the measured field strength as a function of the distance to
the shower axis $R_0$.  The measured field strength as a function of the distance
to the shower axis is presented in \fref{lat} for a typical event.  The
characteristic length $R_0$ for this example is of order of 100~m.

The investigations reveal that there are different types of air showers. Most
of them have a lateral distribution which is characterized by a constant
$R_0\approx100$ to 200~m, like the example shown in \fref{lat}. On the other
hand, there are few showers with relatively flat lateral distributions and
corresponding values for $R_0$ as high as 1000~m or 1200~m.  A closer look
indicates that the steepness of the fall-off $R_0$ depends on the mean distance
to the shower axis $R_{mean}$ and the zenith angle of the showers $\theta$.
The reconstructed scale parameter $R_0$ is plotted as a function of the
relation $(1-\sin\theta)R_{mean}$ in \fref{latpar}.  A correlation between the
two quantities can be recognized.  Large values for $R_0$ are obtained for
showers with large zenith angles and a small average distance between the
shower axis the antennas. But as not all events with small $R_{mean}$ show a
flattening --- the reason is still unclear and further investigations with
larger statistics are required.

\Section{Polarization}

\begin{figure}[t]
 \epsfig{file=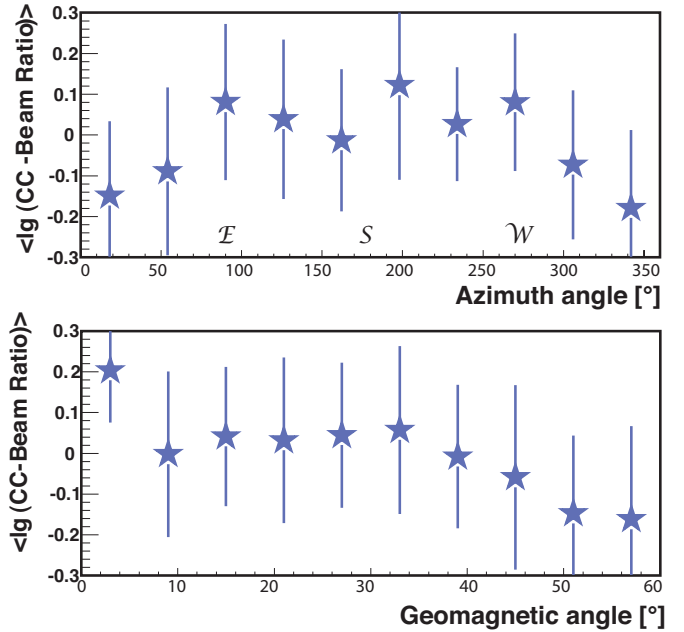,width=\columnwidth}
 \caption{Ratio of the reconstructed cross-correlated beam values of the
	  north-south and east-west polarization components as a function of
	  azimuth angle (top) and the geomagnetic angle (bottom)
          \cite{icrc09-isar}.}
 \label{pol}	  
\end{figure}

An important contribution towards the full understanding of the emission
mechanism of the radio waves in air showers is the investigation of the
polarization characteristics of the measured radio signals \cite{isar-nim2009}.
The 30 LOPES antennas are set up now such that 15 register the east-west
component and 15 antennas measure the north-south component of the electric
field.

The radio emission generated by the geo-synchrotron mechanism is expected to be
highly linearly polarized. The signal is usually present in both polarization
components (E-W and N-S). The signal strength depends on the geomagnetic angle
and thus, on the shower azimuth and zenith angles \cite{huegefalcke}.  The
emission is expected to be polarized perpendicular to the shower axis and the
direction of the geomagnetic field. The polarization characteristics can be
described by a unit polarization vector \vxB, where $\vec{v}$ is the direction
of the shower axis and $\vec{B}$ the direction of the Earth magnetic field
\cite{icrc09-isar}.  Recently, it has been suggested that in addition to the
polarization characteristics also the absolute magnitude of the electric field
is proportional to the Lorentz force \vxB \cite{Ardouin:2009zp}.

\begin{figure}[t]
 \epsfig{file=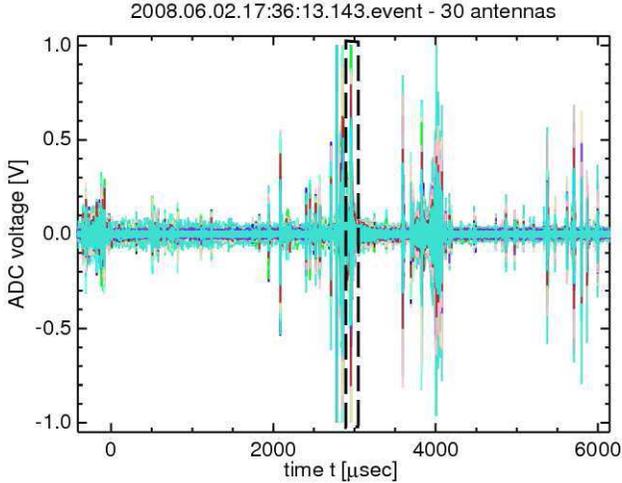,width=\columnwidth}
 \caption{Example of the measured antenna signal as a function of time for a
          thunderstorm event \cite{icrc09-ender}.}
 \label{thunder-ev}	  
\end{figure}

The arrival direction of the cosmic rays on the sky is shown for the registered
events in \fref{polsky}. The color code represents the pulse heights of the
cross-correlated beams, registered in east-west polarization (left) and
north-south polarization (right).  The boxes indicate the directions of the
events with the strongest signals.  The magnetic field in Karlsruhe has a
zenith angle of 25\deg and an azimuth angle of 180\deg.  It can be recognized
that the showers with the strongest signals are registered perpendicular to the
Earth magnetic field: from northern directions in east-west polarization
direction and from west and east for the north-south polarized signal.

To confirm the \vxB behavior of the polarized signals the ratio of the
amplitudes of the cross-correlated beams in east-west and north-south
polarization is investigated \cite{icrc09-isar}. The ratio of the north-south
to east-west amplitudes is depicted in \fref{pol} as a function of the azimuth
angle (top) and the geomagnetic angle (bottom).  Mean values and the spread of
the distributions are shown.  A correlation between the plotted quantities can
be clearly recognized.  The ratio of the north-south divided by the east-west
component of the \vxB vector of the measured showers exhibits the same
behavior as a function of the azimuth angle and geomagnetic angle.  This \vxB
behavior of the measured data is a strong indication for a geomagnetic origin
of the radio emission in air showers.

\Section{Thunderstorm events}

\begin{figure}[t]\centering
 \epsfig{file=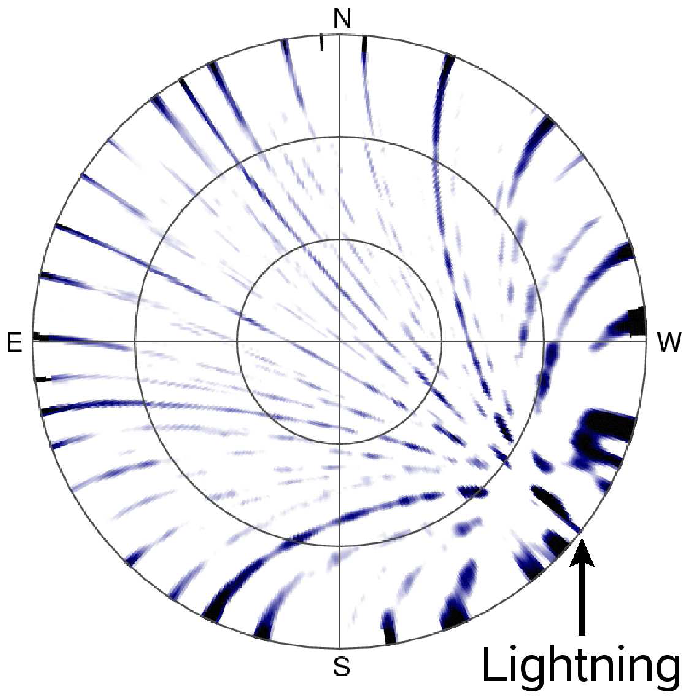,width=0.9\columnwidth}
 \caption{Sky map of the thunderstorm event shown in \fref{thunder-ev}
          \cite{icrc09-ender}.}
 \label{thunder-sky}	  
\end{figure}

The measured field strength of the radio emission of air showers depends on the
(static) electric fields in the atmosphere.  The electric fields inside
thunderstorm clouds, in particular within the convective region can reach peak
values up to 100~kV/m. This leads to additional forces on the electrons and
positrons in the air showers and, consequently, to an acceleration and
deceleration of parts of the electromagnetic shower component (electrons and
positrons, depending on the direction of the electric field).  In turn, this
yields to an amplification or reduction of the radio emission of air showers
during thunderstorms.  Such a behavior has been observed with LOPES
\cite{buitinkthunder}. To obtain reliable information about the detected air
showers from the observations of radio emission requires to monitor the
electric field in the atmosphere and to record the signatures of thunderstorms.
Therefore, the signals form air showers during thunderstorms are studied in
detail \cite{icrc09-ender}.

As an example, the antenna signal as a function of time measured during a
thunderstorm is shown in \fref{thunder-ev}.  A signal from an air shower is
expected at the zero point of the time axis with an amplitude less than 0.1~V.
Strong additional signals caused by lightning can be recognized.  The dashed
lines mark a region with strong lightning signals, which is used to calculate a
sky-map of the cross-correlated beam. The result is presented in
\fref{thunder-sky}. The map shows the whole sky with the zenith in the center
and the horizon at the border.  A strong signal is visible in
south-western direction, marking the discharge region of the lightning.  The
signal extends to the horizon. This indicates that a cloud-ground discharge has
been registered (and not a cloud-cloud discharge).  The grating lobes of the
antenna array extend over the whole sky.

\begin{figure*}[t]
 \epsfig{file=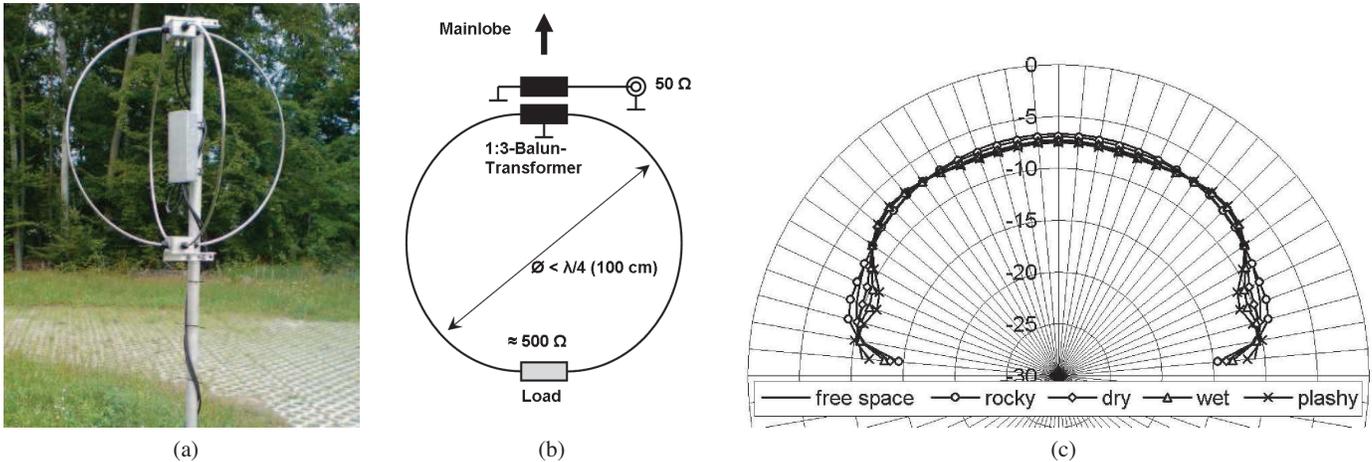,width=\textwidth}
 \caption{Photo of a SALLA antenna and circuit diagram (left).
          E-plane directional diagram (right) \cite{icrc09-kroemer}.}
 \label{salla}	  
\end{figure*}

At present, studies are under way to detect a possible distortion of an event
due to thunderstorm conditions directly from the measured data
\cite{icrc09-ender}. The idea is to find a possible deviation of the
polarization characteristics from the theoretically expected behavior. Such an
indication would point towards a change in the emission process.

\Section{New antenna types and self-trigger algorithms}

An important objective of LOPES is to develop and optimize new antenna types to
measure radio signals from extensive air showers in large installations, such
as the Pierre Auger Observatory, see \sref{outsec}: a sub-project named \Lstar.
Most cosmic-ray radio detectors use (simple) dipole antennas, they are easy to
assemble and are available at relatively low costs.  To avoid uncertainties
related to dipoles new antenna types are developed, such as the logarithmic
periodic dipole antenna (LPDA) with two crossed polarization directions
\cite{gemmekearena}.  These antennas exhibit an almost frequency independent
directional diagram, antenna gain, and impedance.

Another way to design wide-band directional antennas with dimensions much
smaller than the LPDA is given by resistively loaded aperiodic antennas with
internal losses \cite{icrc09-kroemer}.  They also have excellent wide-band
properties as the resistor load dominates in comparison to the capacitive or
inductive reactance.  For a large-scale radio detector a crossed polarized
short aperiodic loaded loop antenna (SALLA) has been developed.  It has a
diameter of only 100~cm (120~cm in the latest version), a weight less than
2~kg, and material costs of about 60~EUR. A photograph of the antenna and the
circuit diagram are depicted in \fref{salla}.  The damping resistor guarantees
a wide bandwidth.  The E-plane directional diagram is shown in \fref{salla}
(right). It features a wide main lobe towards zenith. The 3~dB beam width
amounts to 150\deg and is about 50\deg wider than the directional pattern of
the LPDA.

Ultimate goal is to use these antennas in a self-triggered mode, i.e.\ the data
acquisition is started directly from the detected radio signal
\cite{icrc09-schmidt}.  A trigger algorithm has been developed and realized in
FPGA-hardware. It provides RFI suppression by Fourier transforming the radio
signal to the frequency domain in real time, eliminating mono frequent carriers
and transforming it back to the time domain. Then a threshold is applied and
events are selected which fulfill certain criteria to characterize the pulse
shape.  In a last step, coincidences between three neighboring antennas are
required for a valid event. The real-time feasibility of this trigger mechanism
was proven on prototype hardware.

\Section{Outlook} \label{outsec}

The investigations of radio signals from air showers with LOPES and CODALEMA
are the basis for the application of these technique in large-scale
experiments. The next step is to utilize the radio detection technique in large
arrays, comprising several hundreds of antennas.

LOPES serves as prototype for the detection of radio emission from air showers
in LOFAR \cite{lofar-isvhecri08}.
The Low Frequency Array 
(LOFAR) is a new digital radio observatory, presently under construction in the
Netherlands and in Europe \cite{lofar}. It is
designed as multi-sensor network to assist scientists in the fields of
astronomy, geophysics, and agriculture.  Main focus of the astronomy community
is to observe the radio Universe in the frequency range from 30 to 240~MHz.
An objective of LOFAR is the
detection of radio emission from particle cascades, originating from extremely
high-energy particles from outer space.  Two main lines of research are
followed: (i) the measurement of radio emission from extensive air showers,
generated by interactions of high-energy cosmic rays in the atmosphere 
\cite{icrc09-horneffer} and (ii)
the detection of radio emission of particle cascades in the Moon, originating
from ultra high-energy neutrinos and cosmic rays interacting with the lunar
surface \cite{icrc09-singh}.

More than 40 stations with fields of relatively simple antennas will work
together as digital radio interferometer. The antenna fields are distributed
over several countries in Europe with a dense core in the Netherlands. The
latter will have at least 18 stations in an area measuring roughly
$2\times3$~km$^2$. Each station will comprise 96 low band antennas, simple
inverted V-shaped dipoles (like the LOPES antennas), operating in the frequency
range from 30 to 80~MHz.  Each antenna will have a dipole oriented in
north-south and east-west directions, respectively. In addition, fields
\footnote{The fields comprise 48 antennas in the Dutch stations and 96 in the
European ones.} of high-band antennas will cover the frequency range from 110
to 240~MHz.  For air shower observations the signals from the low band antennas
are digitized and stored in a ring buffer (transient buffer board, TBB).  For
valid triggers the data are send to a central processing facility, based on an
IBM Blue Gene supercomputer.  First data from LOFAR are expected in early 2010.

Goal of LOFAR is to further push the development of radio detection to be a
new, independent way to measure the properties of air showers. Ultimate goal is
to derive information about the primary, shower-inducing particle from the
measurements, such as the particles energy, mass, direction and point of
incidence.
The curvature of the shower front has been investigated
\cite{lafebrecurv,icrc09-lafebre}. It could be shown that the radius of
curvature measured at ground level from the radio observations is related to
the distance to the shower maximum.  The depth of the shower maximum in the
atmosphere \Xmax is an important observable to determine the mass of the
primary particle.  The study indicates a resolution of \Xmax from radio
observations of order of $30-40$~\gcm2.

Another important application is the measurement of radio emission from air
showers at the Pierre Auger Observatory with the Auger Engineering Radio Array
(AERA) \cite{theseprocdallier,icrc09-vdberg}.  AERA is co-located with the
infill array of the Pierre Auger Observatory and in the field of view of the
high-altitude fluorescence telescopes (HEAT). This unique set-up will allow to
register air showers simultaneously with three independent detection methods:
radio waves, fluorescence light, and particle detection in water \Cerenkov
detectors.  AERA comprises about 150 antennas located on an area of about
20~km$^2$ and is designed to cover the energy range from $10^{17}$ to
$10^{19}$~eV.
It has three main science goals:
(i) the thorough investigation of the radio emission from air showers and the analysis of the observed field strength as a function of various air shower parameters, such as energy, angle of incidence, and distance to the shower axis.
(ii) Exploration of the capability of the radio detection technique and to study the feasibility as stand-alone technique to measure the properties 
of air showers.
(iii) To measure the composition of cosmic rays in the energy range from
$10^{17}$ to $10^{19}$~eV.  First data from AERA are expected before Summer
2010.

The radio detection of air showers is a fast growing sub-discipline in
astroparticle physics. With the new experiments starting operation exciting
results are expected in the next few years.


\end{document}